\documentclass{elsarticle}
\usepackage{graphicx}
\usepackage{amssymb}
\usepackage{epstopdf}
\usepackage{bm}
\usepackage{color}

\def\k{{\boldsymbol k}}

\def\x{{\boldsymbol x}}

\def\r{{\boldsymbol r}}
\def\z{{\boldsymbol z}}

\def\v{{\boldsymbol v}}

\def\b{{\boldsymbol b}}

\def\vareps{\mbox{\boldmath$\varepsilon$}}
\def\e{{\boldsymbol e}}

\newcommand{\nn}{\nonumber\\ }
\newcommand{\beq}{\begin{eqnarray}}
\newcommand{\eeq}{\end{eqnarray}}

\journal{Physics Letters B}
\begin{document}


\begin{frontmatter}

\title{Continuous description of fluctuating eccentricities}
\author{Jean-Paul Blaizot$^1$, Wojciech Broniowski$^{1,2,3}$, Jean-Yves Ollitrault$^1$}
\address{$1$ Institut de Physique Th\'eorique, CNRS/URA 2306, F-91191 Gif-sur-Yvette, France\\
$2$ The H. Niewodnicza\'nski Institute of Nuclear Physics, Polish Academy of Sciences, PL-31342 Krak\'ow, Poland\\
$3$ Institute of Physics, Jan Kochanowski University, PL-25406~Kielce, Poland}

\begin{abstract}
We consider the initial energy density in the transverse plane of a high energy
nucleus-nucleus collision as a random field $\rho(\x)$, whose probability
distribution $P[\rho]$, the only ingredient of the present description, encodes
all possible sources of fluctuations. We argue that it is a local Gaussian, with
a short-range 2-point function, and that the fluctuations relevant for the
calculation of the eccentricities that drive the anisotropic flow have small
relative amplitudes. In fact, this 2-point function, together with the average
density, contains all the information needed to calculate the eccentricities
and their variances, and we  derive general model independent expressions for
these quantities. 
The short wavelength fluctuations are shown to play no role in these
calculations, except for a renormalization of the short range part of the
2-point function.  As an illustration, we compare to a commonly used model of
independent sources, and recover the known results of this model.  

\end{abstract}
\begin{keyword}
relativitic heavy-ion collisions \sep harmonic flow \sep correlations

PACS codes here, in the form: 
\end{keyword}

\end{frontmatter}

\section{Introduction}

The fluctuations of the initial energy density (to be denoted $\rho(\x)$
throughout   this paper) in the transverse plane of a heavy ion collision play an
essential role in the dynamics of these collisions. They leave observable traces in particle
distributions after the hydrodynamical
evolution~\cite{Luzum:2013yya}. 
They are for instance responsible for 
elliptic flow
fluctuations~\cite{Alver:2006wh,Andrade:2006yh} 
triangular flow~\cite{Alver:2010gr,ALICE:2011ab,Adare:2011tg,Adamczyk:2013waa}
and higher harmonics~\cite{Chatrchyan:2012wg,ATLAS:2012at},
directed flow near
midrapidity~\cite{ATLAS:2012at,Teaney:2010vd,Luzum:2010fb,Retinskaya:2012ky}, 
and may also explain~\cite{Broniowski:2009fm,Bozek:2012fw}
observed 
transverse momentum 
fluctuations~\cite{Adamova:2003pz,Adams:2003uw,Adler:2003xq,Anticic:2003fd}.
Considerable experimental and theoretical 
efforts are presently devoted to pin down the details of these
fluctuations~\cite{Aamodt:2011by,Gardim:2012im,Heinz:2013bua,CMS:2013bza}
and their various correlations~\cite{ALICE:2011ab,Bhalerao:2011ry,Aad:2014fla}.

It is then a natural
question to try and specify the nature of the information that one  can
extract from measurements of various features of anisotropic flows. 
The initial energy density fluctuations are of several origins. The most
prominent ones are usually attributed to the motion of individual nucleons in the
nuclear wave-functions, and  treated by Glauber Monte Carlo
calculations~\cite{Miller:2007ri,Alver:2008aq,Broniowski:2007ft,
Broniowski:2007nz}. In addition, there are sub-nucleonic  fluctuations, that
reflect the partonic structure of the colliding objects~\cite{Schenke:2012wb}.
In most approaches, such sub-nucleonic fluctuations are added on top of the
geometrical ones, using various
``recipes''~\cite{Dumitru:2012yr,Rybczynski:2013yba}. There is considerable
ambiguity in the whole procedure:  sources, with various
locations~\cite{Bzdak:2013zma}, strengths~\cite{Dumitru:2012yr}, spatial
extents, shapes, etc.,  are added by hand to an already crude description of the
nuclear wave-functions. It would certainly be desirable to use a description
where all irrelevant details  do not stand prominently. 

We find it   useful then to address the question from another angle, with the goal of  obtaining general, model independent, statements about the fluctuations. To achieve this goal, we regard the energy density $\rho(\x)$ in the transverse plane as  a random field, and try to characterize the underlying probability distribution, $P[\rho]$  for finding a given $\rho(\x)$ in a particular event.  This probability distribution is the only ingredient of the distribution, and it encodes all sources of fluctuations, irrespective of their natures. We conjecture that this distribution is a local Gaussian with a short-range 2-point function. That is, we argue  that the fluctuations of the density at different points in the transverse plane are  essentially uncorrelated. Corrections are to be expected in regions where the nuclear density is low, and these corrections will be qualitatively discussed. Furthermore, we also argue that short wavelength fluctuations are irrelevant for the calculations of the 
eccentricities that drive the anisotropic flows, except for a small renormalization of the short range 2-point function. 

We start, in the next section, by deriving general expressions for the
eccentricities and their variances, in terms of the average density and the
2-point function of the probability distribution. The calculation exploits the
fact that the relevant fluctuations have a small amplitude, relative to the
average density. We then provide a simple ansatz for the 2-point function, 
which is  dominated by a short range contribution. We compare results obtained
with this ansatz with those obtained with a model of independent sources, and we
recover known analytic formulas expressing the  eccentricities as products of
geometrical factors by an overall measure of the strength of the fluctuations.  
We then discuss why the Gaussian distribution provides a simple, and presumably
realistic, form for the probability distribution.

\section{Expressions of fluctuation observables in terms of the two-point function}

We   characterize event classes  by the impact parameter $\b$. 
Even though not directly accessible experimentally, the impact parameter is well defined in a high energy collision. For simplicity, in most of this paper, we restrict ourselves to the case of central collisions, i.e. $\b=0$, except for a remark on the general case at the end of this section. We write the energy density  in a given event class as $\rho(\x)=\langle \rho(\x)\rangle+\delta\rho(\x)$, where $\langle \rho(\x)\rangle $ is the average energy density and  $\delta\rho(\x)$ is referred to as the fluctuation. The probability that a given $\rho(\x)$ occurs in the event class considered  is denoted by $P[\rho]$.  

The observables that we wish to calculate  characterize the shape of the fluctuating density  $\rho(\x)$, in terms of its  
moments, commonly referred to as eccentricities \cite{Teaney:2010vd}. These are defined by 
\begin{equation}
\label{defepsn}
\e_n \equiv \int_z\, z^n\, \rho(z).
\end{equation}
Note that $\e_n$ is a vector in the transverse plane (i.e., the plane transverse to the collision axis). In the right hand side of Eq.~(\ref{defepsn}) we use the complex notation to represent vectors in the plane. That is, we allow for a slight abuse of notation and  denote indifferently a vector $\r$  by its components $x,y$, or by the complex number $z=x+iy$. Thus the  density, denoted indifferently by $\rho(\r)$ or $\rho(z)$,  is a real function of $x$ and $y$. Similarly,  we use the short hand $\int_z=\int {\rm d}x{\rm d}y$ for the integration over the transverse  plane.

 The zeroth and first moments are special and require specific definitions:
\beq\label{defepsn1}
\e_0=\int_z |z|^2\rho(z),\qquad  \e_1=\int_z \,z^2\bar z\, \rho(z),
\eeq 
with $\bar z$ denoting the complex conjugate of $z$. The zeroth moment $\e_0$ is
the mean squared radius of the density, while $\e_1$ is a measure of the dipole
moment of the distribution. The particular weight  $z^2\bar z$ in the integral
defining $\e_1$, instead of the more natural one, $z$, is due to the fact that
in a centered coordinate system, to be defined shortly, the dipole moment
vanishes (Eq.~(\ref{centered}) below).

By definition, we call  ``centered'' a  coordinate system where
\beq\label{centered}
\int_z \, z\rho(z)=0.
\eeq
It is only in such a system that the   definitions~(\ref{defepsn}) and (\ref{defepsn1}) above are valid.
In a fixed coordinate system, however, the fluctuating density would be centered
around a random point $z_0$, distinct form the origin,
\beq\label{zzero}
z_0=\frac{\int_z\, z\rho(z)}{\int_z\,\rho(z)},
\eeq
and the definitions above need to be modified accordingly:
\beq
&&\e_0=\int_z |z-z_0|^2\rho(z),\qquad \e_1=\int_z (z-z_0)^2(\bar z-\bar z_0)\rho(z),\nn
&& \e_n =\int_z\,(z-z_0)^n\rho(z).
\eeq
Because $z_0$ is a functional of $\rho$, Eq.~(\ref{zzero}), the averages of the
eccentricities are in general difficult to evaluate. However, simple expressions
can be obtained in the regime of small fluctuations, which is the case of
practical interest. Indeed the calculation of the eccentricities $\e_n$,  with
$n$ small, involves only the lowest (small $\k$) Fourier coefficients
$\delta\rho_\k$  of the fluctuation  $\delta\rho(\r)= \int_\k {\rm
e}^{i\k\cdot\r} \,\delta\rho_\k$   \cite{Teaney:2010vd}. 
This automatically eliminates the rare fluctuations where $\delta\rho(\r)$  can be locally large (spikes). We return to this issue in the next section.

We then assume  that, for the long wavelength fluctuations, $\delta\rho(z)\ll \langle\rho(z)\rangle$, and   choose the
coordinate system such that $\langle\rho(z)\rangle$  is centered; in particular,
at vanishing impact parameter, $\langle\rho(z)\rangle$ has azimuthal symmetry.
The center of mass of $\rho(z)$ is still given by Eq.~(\ref{zzero}) with
$\rho(z)$ in the numerator replaced by $\delta\rho(z)$: it follows therefore
that $z_0$ differs from the origin of the coordinate system by a small amount,
of order $\delta\rho/ \langle\rho\rangle$. A simple calculation then yields, to
linear order in the fluctuation,
\beq
&&\e_0=\int_z \,|z|^2[\langle \rho(z)\rangle+\delta\rho(z)],\qquad \e_1 =\int_z\, [z^2\bar z-2\langle r^2\rangle z] \delta\rho(z),\,\nn
&& \e_n=\int_z\, z^n \delta\rho(z), \,
\eeq
where we have set
\beq\label{msr0}
\langle r^2 \rangle\equiv\frac{ \int_z |z|^2 \langle\rho(z)\rangle}{\int_z\,\langle\rho(z)\rangle},
\eeq
and we have used symmetries of $\langle\rho(z)\rangle $ to eliminate some terms. 
\\

 The anisotropic flow coefficients $\v_n$ that are experimentally measured, are not directly related to the $\e_n$'s, but are rather proportional to  the
dimensionless ratios \cite{Alver:2006wh,Teaney:2010vd} defined, in a centered system, by 
\beq
\label{defepsn2}
\vareps_n \equiv \frac{\int_z \, z^n \rho(z)}{\int_z \, |z|^n\rho(z)},\qquad
 \vareps_1=\frac{\int_z \,z^2\bar z\,\rho(z)}{\int_z \, |z|^3\rho(z)}.
\eeq
 It has been shown indeed that the relation $\v_n\propto\vareps_n$, is well satisfied in ideal 
hydrodynamics~\cite{Holopainen:2010gz,Qiu:2011iv,Gardim:2011xv}, and even better so in viscous hydrodynamics ~\cite{Niemi:2012aj}.
Note that with the sign convention chosen in Eq.~(\ref{defepsn2})  (which differs from that in Ref.~\cite{Teaney:2010vd}) the response coefficients $v_n/\vareps_n$ are {\it negative\/}.
Similarly, we define  $\vareps_0$ by dividing $\e_0$ by the total energy: 
\beq
\label{defepsn0}
  \vareps_0=\frac{\int_z |z|^2\,\rho(z)}{\int_z \, \rho(z)}.
\eeq  
This (dimensionful) quantity represents the  mean square radius of the distribution in an individual event. It is distinct from (\ref{msr0}) which involves the average density. 

 Expanding the scaled moments (\ref{defepsn2}), (\ref{defepsn0}) in powers of the fluctuation, we obtain,  to leading order,
\beq
\label{expeps0}
\vareps_0
=\langle r^2\rangle+\frac{\int_z \delta\rho(z) (|z|^2-\langle r^2\rangle)}{\int_z \langle\rho(z)\rangle},
\eeq
and 
\beq\label{centering0}
\label{expepsn}
\vareps_n=\frac{\int_z\,z^n\,\delta\rho(z) }{\int_z \,|z|^n \langle\rho(z) \rangle},\qquad 
\vareps_1=\frac{\int_z\, [z^2\bar z-2z\langle r^2\rangle]\,\delta\rho(z)}{\int_z\,|z|^3\langle\rho(z) \rangle }.
\eeq
 Note that,  at this order, only $\vareps_0$ contains a contribution unrelated
to fluctuations, all eccentricities $\vareps_n$ with $n\ge 1$, being entirely
due to fluctuations for central collisions (the numerators of
Eq.~(\ref{centering0}) are proportional to $\delta\rho$, the contributions of
$\langle\rho\rangle$ being zero for symmetry reasons). 
 
The situation is different at non zero impact parameter. The  calculations of
the eccentricities for finite $\b$ are straightforward extensions of those just
presented for $\b=0$, and lead to corrections to the formulas above. Just as an
illustration, let us indicate the expressions of $\vareps_2$ and $\vareps_3$ at 
finite $\b$ and to leading order in the fluctuation:
\beq\label{centering1}
\label{expepsnb}
&&\vareps_2=\frac{\langle z^2\rangle}{\langle r^2\rangle}+ \frac{\int_z\,[z^2-|z|^2\langle z^2\rangle/\langle r^2\rangle]\,\delta\rho(z) }{\int_z \,|z|^2 \langle\rho(z) \rangle},\qquad\langle z^2 \rangle\equiv\frac{ \int_z z^2 \langle\rho(z)\rangle}{\int_z\,\langle\rho(z)\rangle},\nn
&&\vareps_3=\frac{\int_z\, [z^3-3 \langle z^2\rangle z] \,\delta\rho(z) }{\int_z \,|z|^3 \langle\rho(z) \rangle}, 
\eeq
where the averages are taken here for a non vanishing  impact parameter.
Now, $\vareps_2$  also contains a correction independent of the fluctuation,
whose physical interpretation is clear: the contribution $\langle
z^2\rangle/\langle r^2\rangle$ represents the usual average eccentricity related
to the ``almond'' shape of the collision zone. The odd harmonics, on the other
hand, such as $\vareps_3$, remain entirely due to fluctuations.

 We return now to the case $\b=0$. Since the eccentricities are linear in $\delta\rho$, their variances can be easily determined in terms of the two-point function of the probability distribution,  defined as 
  \beq\label{defS}
S(z_1,z_2)  \equiv \langle\delta\rho(z_1)\delta\rho(z_2)\rangle=\langle\rho(z_1)\rho(z_2)\rangle-\langle\rho(z_1)\rangle\langle\rho(z_2)\rangle.
\eeq
Equation (\ref{expeps0}) yields 
 \beq\label{excvariances0}
\langle \Delta\vareps_0^2\rangle=\frac{\int_{z_1z_2} \,(|z_1|^2-\langle r^2\rangle])(|z_2|^2-\langle r^2\rangle) \,S(z_1,z_2)}{\left(\int_z\,\langle\rho(z)\rangle\right)^2}
\eeq
  while the mean square (ms) eccentricities are given by 
\beq\label{excvariances1}
\langle \Delta\vareps_n^2\rangle=\langle \vareps_n\bar{\vareps}_n\rangle& =&\frac{\int_{z_1z_2} \,z_1^n\,\bar z_2^n\, S(z_1,z_2)}{\left( \int_z\, |z|^n\langle\rho(z) \rangle  \right)^2 },\nonumber\\  \,&\, &\, \nonumber \\
\langle \Delta \vareps_1^2\rangle=\langle  \vareps_1\bar{\vareps}_1\rangle&=&\frac{\int_{z_1z_2} \, (|z_1|^2-2\langle r^2\rangle) (|z_2|^2-2\langle r^2\rangle)\, z_1\bar z_2  \,S(z_1,z_2)}{\left( \int_z\, |z|^3\langle\rho(z) \rangle  \right)^2 }.\nn
\eeq

These formulas are general. They show that,  in the regime of small
fluctuations, the eccentricities and their variances are determined entirely by 
the average density and the two-point function of the probability distribution
function $P[\rho]$.

\section{Simple ansatz for  the two-point function}

At this point, it is instructive to consider a simple ansatz for the two-point function. This ansatz is motivated by considerations that will be discussed in the next section, as well as by the connection that  it allows with simple models that have already been used to calculate the eccentricities.  We assume that the two-point function is of the form
\beq\label{ansatzS12}
S(z_1,z_2)=A(z_1)\delta(z_1-z_2)+B f(z_1)f(z_2),\eeq
where $\delta(z_1-z_2)=\delta(x_1-x_2)\delta(y_1-y_2)$, and the dependence of $A$ on $z$ is through its dependence on
 $\langle\rho(z)\rangle$, in other words, $A$ is a function of $\langle \rho(z)\rangle$.  In Eq.~(\ref{ansatzS12}),  $B$ is a constant and $f$ some function, whose general determination is delayed till next section.  However, an important practical case, as we shall see later, is $f(z)=\langle \rho(z) \rangle$. In this case the  term proportional to $B$ in Eq.~(\ref{ansatzS12}) does not contribute to the variances (\ref{excvariances0}) and (\ref{excvariances1}).  
Inserting ansatz (\ref{ansatzS12}) into Eqs.~(\ref{excvariances0}) and
(\ref{excvariances1})
yields:
\beq
&&\langle \Delta\vareps_0^2\rangle=\frac{\int_z\, A(z) (|z|^2-\langle r^2\rangle)^2}{\left( \int_z\, \langle\rho(z)\rangle\right)^2},\qquad\langle \vareps_1\bar{\vareps}_1\rangle=\frac{\int_z (|z|^2-2\langle r^2\rangle)^2 |z|^2  A(z)}{\left( \int_z\langle\rho(z) |z|^3\rangle  \right)^2 }, \nn
&&\langle \vareps_n\bar{\vareps}_n\rangle=\frac{\int_z\, |z|^{2n}\, A(z)}{\left( \int_z \langle\rho(z) |z|^n\rangle  \right)^2 }.
\eeq
The function $A(z)$ is unknown at this stage. However, the success of the model
of independent sources, that we shall introduce shortly, suggests that a linear
dependence on the average density may be a good approximation. For this
particular choice, namely $A(z)=A\left<\rho(z)\right>$,  with $A$ constant, the
formulas  above simplify further:
\beq\label{Deltaeps}
&&\langle \Delta\vareps_0^2\rangle=\frac{A}{ \int_z \langle\rho(z)\rangle}\,\left[ \langle r^4\rangle -\langle r^2\rangle^2\right],\nn
&&\langle \vareps_1\bar{\vareps}_1\rangle=\frac{A}{\int_z \langle\rho(z)\rangle}\;\frac{\langle r^6\rangle-4\langle r^4\rangle\langle r^2\rangle+4\langle r^2\rangle^3}{ \langle r^3\rangle^2},\nn
&& \langle \vareps_n\bar{\vareps}_n\rangle=\frac{A}{\int_z \langle\rho(z)\rangle}\;\frac{\langle r^{2n}\rangle}{ \langle r^n\rangle^2}.
\eeq
These formulas give the eccentricities as products of a ``geometrical'' factor
that depends on the specific observable, and a dimensionless coefficient ${A}/{
\int_z \langle\rho(z)\rangle}$, independent of the observable, that
characterizes the magnitude of the local fluctuations of the energy density. At
this point, it cannot be determined otherwise than by fitting the eccentricities
to experimental data. However, further insight can be gained by considering the
model of independent sources that we just alluded to.

This model can be viewed as a parametrization of the continuous energy density $\rho(z)$  is in terms of ``sources''  \cite{Alver:2006wh,Holopainen:2010gz,Bhalerao:2011bp}. 
One writes, typically
\beq
\rho(z)=E_0\sum_{i=1}^{N_s} \delta(z-z_i),\qquad \int_z\rho(z)=N_s E_0,
\eeq
where $z_i$ denotes the random location of a source,  $N_s$ the number of sources, and $E_0$ the energy of a source, taken here to be constant for simplicity.  If one considers a class of events with a given transverse energy, then $N_s$ is also a constant. In this case, the fluctuations are entirely due to the random positions of the sources in individual events. If  the positions of these sources are uncorrelated, the 2-point function is proportional to that of independent particles in the plane. It reads
\beq\label{independentS12}
S(z_1,z_2)=E_0\langle \rho(z_1)\rangle\delta(z_1-z_2) -\frac{1}{N_s}\langle \rho(z_1)\rangle\langle \rho(z_2)\rangle.
\eeq
This is a particular case of Eq.~(\ref{ansatzS12}), with $A(z)=A\langle
\rho(z)\rangle$, $A=E_0$,  and $f(z)=\langle\rho(z)\rangle$, $B=1/N_s$. Note
that  the  second term in Eq.~(\ref{independentS12})  ensures that fluctuations
conserve the total number of sources, such that  in particular $\int_{z_1 z_2}
S(z_1,z_2)=0$. This feature will be elaborated upon at the end of the next
section. Here we simply notice that  it affects only moderately the local
fluctuations. To see that, let us  assume for simplicity that the average
density is constant on the transverse plane, and consider a small area
$\sigma\ll \Sigma$, where $\Sigma$ is the total area of the collision zone. From
Eq.~(\ref{independentS12}) one then easily deduces
\beq
\langle \delta\rho^2\rangle=\frac{E_0}{\sigma} \langle \rho\rangle\left(1-\frac{\sigma}{\Sigma}   \right),
\eeq
where $\langle\rho\rangle=E_0N_s/\Sigma$.  
The last term in this expression, which originates from the second term in Eq.~(\ref{independentS12}),  is indeed negligible if $\sigma\ll \Sigma$. In this case, one recovers, for  the number of sources in the area $\sigma$, $N_\sigma=\rho\sigma/E_0$, the variance characteristic of Poisson statistics, $\langle \delta N_\sigma^2\rangle=\langle N_\sigma\rangle$. By increasing $\sigma$, one probes fluctuations of $N_\sigma$ over regions that are comparable to the total transverse area of the collision zone, that is, one looks at long wavelength density fluctuations. The distribution of $N_\sigma$ becomes then a Gaussian, with a width $\propto\sqrt{\langle N_\sigma\rangle}$. The relative fluctuations are then small, with $\sqrt{\langle \delta\rho_\sigma^2\rangle/\langle\rho_\sigma\rangle^2}=1/\langle N_\sigma\rangle\ll 1$, which confirms the validity of the expansion used in Sect.~2, based on the relative smallness of the amplitude of the long wavelength fluctuations. 

More generally, we may characterize the strength of the fluctuations by the following ratio
\beq\label{fluctcharacter}
\frac{\int_{z_1 z_2} A(z_1)\delta(z_1-z_2)}{\left(  \int_z\langle  \rho(z)\rangle \right)^2}=\frac{E_0}{\int_z\langle  \rho(z)\rangle}=\frac{1}{N_s},
\eeq
where, in the right hand side, we have used the 2-point function (\ref{independentS12}) of the source model. Thus, in this model, the dimensionless parameter that characterizes the magnitude of the fluctuations is the total number of sources: This is is natural since, in this model, the fluctuations of the density  are entirely determined by those of  the locations of the sources. By substituting $1/N_s$ for the factor $A/\int_z\langle\rho(z)\rangle$ in Eqs.~(\ref{Deltaeps}), one recovers the expressions  derived in  Refs.~\cite{Bhalerao:2006tp,Bhalerao:2011bp} (see also \cite{Alver:2008zza}). Note the ratio in Eq.~(\ref{fluctcharacter}), since it involves integrals over the transverse plane,  provides a global characterization of the fluctuations. The numerator in Eq.~(\ref{fluctcharacter}) is the integral of the short range part of the correlation function $S(z_1,z_2)$ in Eq.~(\ref{independentS12}). This is obviously equal to the integral of the second term, since, as already noticed,  $\int_{z_1z_2}S(
z_1,z_2)=0$.

The source model is much inspired by the participant picture of
the nucleus-nucleus collisions, and the empirical proportionality between 
the transverse energy and the number of participants $N_{\rm part}$ 
\cite{Miller:2007ri}. Indeed, a comparison of the model with the Glauber Monte
Carlo calculations reveals that the number of sources is of the order of that of
the number of participants $N_{\rm part}$ (roughly one
half~\cite{Bhalerao:2011bp}). However, in the present context,  the proper
interpretation of this model is that of a parametrization of a continuous field,
the fluctuating density $\rho(z)$. 
In particular, there is a priori no reason to correlate rigidly the positions of the sources to those of the participants: in the model discussed above, these sources are located randomly in the collision area.\\

It should be emphasized  that $A(z)$ need not be a simple linear function of
$\langle \rho(z) \rangle$, as we have assumed in the second part of this section
(and as it emerges naturally in the model of independent sources). In fact we
have evidence that in nucleus-nucleus collisions, deviations from this simple
behavior occur near the surface where the average density is small. For
instance, Glauber Monte Carlo calculations suggest a specific type of
correlations, that we  dubbed ``twin correlations, which lead to a
renormalization of the short range part of the two-point
function~\cite{longpaper}.  For the sake of illustration, we take these into
account by assuming that $\langle \rho(r)\rangle =\rho$ is uniform within a disk of
unit radius, such that $\langle r^{2n}\rangle =R^{2n}/(n+1)$ (with $r=|z|$), and that
$A$ picks up an extra contribution near the surface, that is, we set
$A(r)=\rho(1+\alpha \delta(r/R-1))$, with $\alpha $ a small parameter. Then, a simple
calculation reveals that the variances  $\langle|\vareps_1|^2\rangle$, $\langle|\vareps_2|^2\rangle$, $\langle|\vareps_3|^2\rangle$ are corrected respectively by the factors 1, $1+3\alpha$, $1+4\alpha$. The same ordering between $\vareps_2$ and $\vareps_3$ is observed in Monte-Carlo Glauber calculations~\cite{longpaper}.

\section{The probability distribution \label{sec:prob}}

The simplest, least biased, picture for the fluctuations of the energy density in the transverse plane is that of independent fluctuations, with a  probability distribution of the form 
\beq\label{distributionP}
P[\delta\rho]\propto \exp\left \{ -\frac{1}{2}\int_{z_1,z_2}\, \delta\rho(z_1)K(z_1,z_2) \delta\rho(z_2)\right\}.
\eeq
In this expression, the exponent may be seen as the analog of a Landau-Ginsburg free energy \cite{LLStatPhys}. 
The kernel  $K(z_1,z_2)$, which depends on the average local density, is the functional inverse of the  two-point function $S(z_1,z_2)$:
 \beq\label{ansatz2}
 \int_z K(z_1,z) S(z,z_2)=\delta(z_1-z_2),\qquad \langle\delta\rho(z_1) \delta\rho(z_2) \rangle=S(z_1,z_2), 
 \eeq
and, as we shall argue soon, for all practical purposes,  $K(z_1,z_2)\propto\delta(z_1-z_2)$, and therefore also $S(z_1,z_2)\propto\delta(z_1-z_2)$. The distribution (\ref{distributionP}) is then a local functional of $\delta\rho(z)$, describing  uncorrelated fluctuations. 

Aside from the fact that there is little evidence for deviations from this simple ansatz\footnote{Note that a Gaussian distribution for the local  fluctuations of the density does not preclude the existence of high order cumulants for observables such as eccentricities.}, there are microscopic arguments suggesting that energy density fluctuations are correlated only over short distances: At high energy, the processes dominating energy density production are semi-hard processes, and these take place locally, depositing energy over regions of sub-nucleonic sizes. In the Color Glass picture, for instance,  the typical transverse size is of the order of the inverse of the saturation momentum $Q_s$~\cite{Schenke:2012wb}. This is also the distance over which fluctuations are correlated~\cite{Muller:2011bb}. Correlations at the nucleon scale,  inherited from the positions of the nucleons in each nuclei, are presumably not important: if one observes two fluctuations within a sub-nucleonic distance, it is likely that 
they originate from two nucleons that are separated longitudinally, and are therefore uncorrelated.\footnote{ This argument has to be corrected as we move towards the surface of the collision zone, where the density is low, and  twin correlations develop \cite{longpaper}. Recall, however, that these correlation simply renormalize the strength of the short range part of the correlation function (see the discussion at the end of Sect.~3).}\\

There is however another, perhaps stronger, argument in favor of the ansatz (\ref{distributionP}), with a short range kernel: short wavelength fluctuations are mostly irrelevant in the calculation of the eccentricities. In the extreme coarse-graining that is involved in such calculations, where only the smallest Fourier modes are retained, all what the short wavelength fluctuations do is  renormalize the coefficient of the delta function contribution in $S(z_1,z_2)$, that is, the function $A(z)$ in Eq.~(\ref{ansatzS12}). Since this is an important point, let us illustrate it by an elementary calculation. Let us return to the calculation of the eccentricities performed in the previous sections, and 
 smear
the delta function $\delta(z_1-z_2)$  by a smooth, normalized, function $h(z_1-z_1)$ peaked at small values of $|z_1-z_2|$
 (typically a Gaussian of width $\sigma=0.4$~fm~\cite{Bozek:2012fw,Holopainen:2010gz}).
That is, let us set
\beq
S(z_1,z_2)=A(z) h(s),\qquad z\equiv\frac{z_1+z_2}{2},\quad s\equiv z_1-z_2,
\eeq
and consider the calculation of $\langle|\vareps_n|^2\rangle$. After  performing the appropriate expansion that exploits the fact that in the integration $|s|\ll |z|$, we get 
\beq\label{moments}
\int_{z_1z_2} z_1^n \bar z_2^n\,S(z_1,z_2)=m_n\left[  1-\frac{m_{n-1}}{m_n} \frac{n^2}{4}\int_s |s|^2 h(s) \right]\equiv m'_n, 
\eeq
where  $m_n\equiv \int_z A(z) |z|^{2n}$ is the value of the integral when $h(s)=\delta(s)$. 
The finite range of $h$ induces corrections that, for small $n$ (in practice $n\le 6)$, are small, of order $[\int_s |s|^2 h(s)]/\Sigma$, with $\Sigma$ the area of the collision zone. Furthermore,  these corrections can be absorbed in a renormalization of the function $A(z)$, $A(z)\longrightarrow A'(z)$, with the $n^{\rm th}$ moment of $A'(z)$ equal to  $m'_n$ in Eq.~(\ref{moments})  above. Thus the result is completely insensitive to the smearing of the short range  2-point function, provided the range of this smearing stays small enough. These considerations also indicate that the measurements of the eccentricities cannot yield direct information on the  ``granularity'' of the
initial energy density~\cite{Schenke:2012wb,ColemanSmith:2012ka,Floerchinger:2013vua}, nor on the direct microscopic mechanisms responsible for the fluctuations: the eccentricities are sensitive only to the overall strength of the fluctuations, not to the details of what happens at short distance scales.

Finally, we return to our ansatz  for the 2-point function and comment on  the
second term  in Eq.~(\ref{ansatzS12}). This term reflects generically the long
range correlations that originate from global constraints, such as for instance
energy conservation, or recentering. Such constraints can be expressed
generically
in the form
\beq
\int_z \delta\rho(z)\phi(z)=0.
\eeq
Choosing for instance $\phi(z)=1$ enforces $\int_z\delta\rho(z)=0$ in each event
(which corresponds to a selection of events with a given $E_T$), $\phi(z)=z$ 
enforces the centering of the fluctuations\footnote{Note however that the
procedure of modifying the probability distribution in order to deal only with
centered density fluctuations is not equivalent to that used in Sect. 2 where we
shift explicitly the observables by the center of mass $z_0$ of the fluctuation,
and retain all fluctuations in the calculation. The latter procedure is the
correct one to use in the present context.}, $\int_zz\delta\rho(z)=0$, etc.
Implementing such linear constraints on the probability distribution can be
easily done  by using the standard techniques of generating functionals and
Lagrange multipliers.
 The resulting distribution is still a Gaussian, but the two-point function takes now the form
\beq
\langle \delta\rho(z_1)\delta\rho(z_2)\rangle=A(z_1,z_2)-\frac{\left[\int_{\z_2}\,A(z_1,z_2)\phi(z_2)\right] \left[\int_{z_1}\, \phi(z_1) A(z_1,z_2)\right]}{\int_{z_1z_2} \phi(z_1)A(z_1,z_2)\phi(z_2)},
\eeq
where $A(z_1,z_2)$ denotes here the inverse of $K(z_1,z_2)$, i.e., the two-point
function without the constraint. This equation is indeed of the form
(\ref{ansatzS12}), with $f(z)\propto \int_{z'} A(z,z') \phi(z')$. For the
constraint on $E_T$ where $\phi(z)=1$, and the choice
$A(z)=E_0\langle\rho(z)\rangle$, with $\int_z \langle\rho(z)\rangle=E_0 N_s$, we
recover Eq.~(\ref{independentS12}) of the main text. 

\section{Summary}

In summary, we have presented the first steps for a field theoretical description of the initial  fluctuations of the energy density in the transverse plane of a high energy nucleus-nucleus collision. This description, free of irrelevant, model dependent details, rests entirely on the probability distribution for the random field $\rho(\x)$, which we have argued is a Gaussian with a short range 2-point function. We have obtained general expressions for the eccentricities that drive the anisotropic flows in terms of this 2-point function. We have argued that short wavelength fluctuations play no role in this calculation aside from renormalizing slightly the short range 2-point function. If, as we have argued, the basic correlations occur at the sub-nucleonic level, with no other scale playing a major role all the way to the nuclear scale, the present description would hold for systems of all sizes, i.e., from AA to pA, till, why not, to large multiplicity pp.

\section*{Acknowledgements}
This work is 
supported by the European Research Council under the
Advanced Investigator Grant ERC-AD-267258, and by the Polish National Science Centre, grants DEC-2012/06/A/ST2/00390 and DEC-2011/01/D/ST2/00772.    
WB was supported in part by the Polish National
Science Center, grant DEC-2012/06/A/ST2/00390.


\end{document}